# Experimental investigation of electronic interactions in collapsed and uncollapsed $LaFe_2As_2$ phases

Ilaria Pallecchi [1], Elena Stellino [2], Paolo Postorino [3], Akira Iyo [4], Hiraku Ogino [4], Marco Affronte [5], Marina Putti [6,1]

[1] CNR-SPIN, c/o Dipartimento di Fisica, via Dodecaneso 33, 16146 Genova, Italy
[2] Department of Physics and Geology, University of Perugia, Via A. Pascoli, 06123 Perugia, Italy
[3] Department of Physics, Sapienza University of Rome, P.le A. Moro 5, 00185 Rome, Italy
[4] National Institute of Advanced Industrial Science and Technology (AIST), 1-1-1 Umezono, Tsukuba, Ibaraki 305-8568, Japan
[5] Dipartimento di Scienze Fisiche, Informatiche e Matematiche, Università degli Studi di Modena e Reggio Emilia, and CNR-NANO, via G. Campi, 213/A, 41100 Modena, Italy
[6] Università di Genova, Dipartimento di Fisica, via Dodecaneso 33, 16146 Genova, Italy

**Abstract**
The iron-based pnictide $LaFe_2As_2$ is not superconducting as-synthesized, but it becomes such below $T_c \sim 12$ K upon annealing, as a consequence of a structural transition from a phase with collapsed tetragonal crystal structure to an uncollapsed phase. In this work, we carry out specific heat, Raman spectroscopy and normal state electric and thermoelectric transport measurements in the collapsed and uncollapsed $LaFe_2As_2$ phases to gain insight into the electron interactions and their possible role in the superconducting pairing mechanism. Despite clear features of strong electron-phonon coupling observed in both phases, neither the low energy phonon spectra nor the electron-phonon coupling show significant differences between the two phases. Conversely, the Sommerfield constants are significantly different in the two phases, pointing to much higher electron correlation in the superconducting uncollapsed phase and confirming theoretical studies.

## 1. Introduction

Since the first synthesized samples [1], the stoichiometric compound $LaFe_2As_2$ has puzzled the scientific community for becoming superconducting below $T_c \sim 12$ K. Indeed, compared to the undoped pnictide parent compounds $REFe_2As_2$ (RE = rare earth), $RE^{2+}$ is replaced by $La^{3+}$, resulting in a nominal doping of 0.5 electrons/Fe, which would place it in the dramatically overdoped regime, where $T_c$ commonly vanishes in unconventional superconductors with a dome-shaped phase diagram. Actually, bulk superconductivity appears only in the so-called tetragonal "uncollapsed" phase of $LaFe_2As_2$, having elongated c-axis, whereas the "collapsed" counterpart, having shorter c-axis, is not superconducting [1]. Annealing at 500°C of as-synthesized samples drives the structural transition from the latter to the former phase, with significant changes in structural parameters, electronic band structure, electronic properties, and Fe magnetic moments [2,3,4,5].

Ab initio calculations of structural, magnetic, and electronic properties of $LaFe_2As_2$ [6] evidenced the roles played by Fe $d_{xy}$ and $d_z$ orbitals, as well as by the hybridization of the $5d_{xy}$ and As $4p$ orbitals, in determining the electronic properties and superconducting ground state. Orbital resolved calculations of Pauli susceptibility showed that Fe $3d$ states, as well as As $4p$ and La $5d$ states, all contribute significantly to instabilities around $(\pi,\pi,\pi)$ in uncollapsed $LaFe_2As_2$, providing a possible paring glue [7]. Correlation and enhanced scattering in the $d_{xy}$ band, resulting in intense low energy spin fluctuations, were indicated as key ingredient for the unconventional Cooper pair formation [4,5]. The role of strong correlations was confirmed by theoretical calculations that were able to reproduce the experimental Sommerfeld specific heat coefficient $\gamma$ of $LaFe_2As_2$, using the same interaction values which captures the evolution of $\gamma$ of electron- and hole-doped $BaFe_2As_2$ for a large number of doping values [8]. Remarkably, the $\gamma$ value of $LaFe_2As_2$ is twice as much the value predicted by uncorrelated band theory [8].

On the experimental side, the combined analysis of normal state of magneto- and thermo-electric transport and specific heat in the uncollapsed phase evidenced a high effective mass and distinctive features of a strong electron-phonon coupling, typical of conventional superconductors rather than unconventional ones

[9]. Although the extracted transport electron-phonon coupling $\lambda_{tr}$ would be associated to a negligibly small $T_c$ according to the Bardeen-Cooper-Schrieffer evaluation, coupling with phonons could play a role, not only in the normal state, but also in the superconducting mechanisms of uncollapsed LaFe$_2$As$_2$, possibly interplaying with other pairing mechanisms. To better investigate this issue, a comparison of phonon spectra of collapsed and uncollapsed phases could be revealing.

In this work, we present specific heat, Raman spectroscopy, normal state electric and thermoelectric transport studies of collapsed and uncollapsed phases of LaFe$_2$As$_2$, with the specific purpose of detecting differences in phonon spectra, electron-phonon coupling, and correlations, possibly playing a role in the superconducting pairing mechanism.

## 2. Experimental

Collapsed and uncollapsed LaFe$_2$As$_2$ polycrystalline samples were synthesized using a high-pressure and high-temperature synthesis method with subsequent annealing at 500°C as described in the ref. [1].

Powder X-ray diffraction (XRD) patterns were measured at room temperature using a diffractometer (Rigaku, Ultima IV) with Cu K$\alpha$ radiation (wavelength of 1.5405 Å).

Resistivity, magnetotransport and Seebeck measurements were carried out in a Physical Property Measurement System (PPMS) by Quantum Design, in applied magnetic fields up to 9T and at temperatures down to 5 K.

Raman measurements were carried out at room temperature by a Horiba LabRAM HR Evolution microspectrometer, operating in backscattering geometry, coupled with a 600 grooves/mm diffraction grating and a Peltier-cooled charge-coupled device (CCD) detector. In this configuration, it was possible to reach a spectral resolution better than 3 cm$^{-1}$. A 632.8 nm He–Ne laser was employed as a light source and focused on the sample by a microscope equipped with a 100x objective, to obtain a ~1µm diameter spot on the measured surface [10]. Polarization-dependent Raman spectra were collected using a $\lambda/2$ waveplate placed at the microscope entrance so that both incident and scattered radiation can be rotated by a chosen angle $\theta$. A polarization analyser was then placed before the diffraction grating to select the component of the scattered radiation with vertical polarization [10]. For further details see the Supplementary Material file.

Heat capacity measurements were performed in a PPMS system of the Quantum Design by the relaxation method with typical temperature pulse of $\Delta T \approx 1\%$ with respect to the bath temperature.

## 3. Results and analysis
### 3.1 Structural characterization

Phase identification of collapsed and uncollapsed phases was done through the analysis of X-rays diffraction patterns, and the results, shown in Figure 1, are consistent with previous reports [1]. In particular, small amounts of LaFeAsO impurity phase was detected in the collapsed sample and small amounts of LaFeAsO, LaAs, and FeAs impurity phases in the uncollapsed sample.

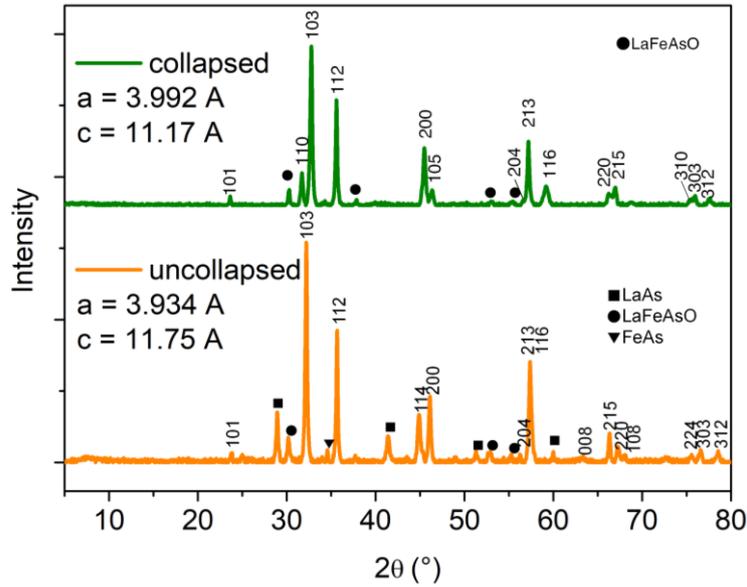

**Figure 1:** X-rays patterns of collapsed and uncollapsed samples. For each pattern, Bragg indices of the main phase and peaks of secondary phases are indicated. The peaks of the LaFeAsO impurity phase can be also attributed to superconducting LaFeAsO$_{1-y}$ or LaFeAs(O,H) phases.

LaAs and FeAs are metallic phases and LaFeAsO, if oxygen deficient or if incorporating H may become superconducting. The contribution of these impurity phases may affect resistivity, which is representative of the lowest resistive percolative path rather than of the bulk, hindering a reliable quantitative analysis. In this work we thus focus the comparison of resistivity curves of collapsed and uncollapsed phases on a qualitative basis and we do not discuss either those aspects that depend on the resistivity magnitude, which may be affected by the impurity phases, as well as by uncertain geometric factors related to granularity. On the other hand, Seebeck effect does not depend on geometric factors and it is more representative of the main phase. Raman measurements, probing locally the material on the scale of one µm, should represent the main phase. Specific heat, which is a bulk property, is affected by the presence of impurity phases proportionally to their mass content, hence the minor contribution of spurious phases can be estimated, if necessary. In the low temperature regime, specific heat allows to identify superconducting phases. Our measurements, presented in the following section 3.4, evidence a bulk superconducting transition around 12 K only in the uncollapsed sample, which is consistent with previous reports [1], where appearance of bulk superconductivity around 12 K and absence of bulk superconductivity were evidenced in the uncollapsed and collapsed phases, respectively.

**3.2 Magnetotransport and thermoelectric properties**

In the upper panel of Figure 2, the resistivity ρ curves of collapsed and uncollapsed samples are shown. In the case of the uncollapsed sample, the superconducting transition at 12 K is seen. For the non-superconducting collapsed sample, the resistivity curve exhibits a downturn with onset at 30K, which is due to the presence of superconducting LaFeAsO$_{1-y}$ or LaFeAs(O,H) impurity phases formed under high pressure, as detected by X-rays diffraction. The most striking difference that is observed form Figure 2 is the much weaker temperature dependence of the resistivity of the collapsed phase. The residual resistivity ratio (RRR), which is evaluated as the ratio of room-temperature to low-temperature resistivities (choosing a low-temperature resistivity ρ(T=50K)), is more than 3 times smaller in the collapsed phase as compared to the uncollapsed phase. This smaller RRR may be related to a higher degree of disorder in this sample, which enhances electron scattering by defects at low temperatures, whereas the higher order in the uncollapsed phase may be related to the annealing step carried out in the preparation of this phase.

It could be argued that the absence of superconductivity in the collapsed phase may be due to the much larger atomic disorder in this sample, recovered in the uncollapsed sample after annealing. Indeed, it is expected that non magnetic atomic disorder has a pairbreaking effect in sign-changing s-wave superconductivity, the most likely scenario for the gap symmetry in iron-based superconductors, and the T$_c$ suppression crucially depends on the ratio of interband to intraband scattering rates. If we compare the $T_c$ suppression observed in 122 pnictide single crystals, where controlled and increasing amounts of atomic disorder was introduced by electron irradiation with a small recoil energy [11,12,13,14], we observe that, in correspondence of residual resistivity enhancements $\Delta\rho$ around few tens of $\mu\Omega$ cm, $T_c$ values decrease by up to ~50% (30% if only optimally doped samples are considered), while a full suppression of superconductivity was not observed. Indeed, iron-based superconductors are generally more robust to disorder than cuprate superconductors, for which electron irradiation produces a complete suppression of superconductivity [15]. The residual resistivity enhancement from the uncollapsed to the collapsed phase cannot be estimated, due to the polycrystalline nature and the related inevitable contributions of grain boundaries and porosity. However, setting an upper limit for the residual resistivity enhancement around 20 $\mu\Omega$ cm, we could expect a maximum $T_c$ suppression by disorder of 20%-30% in the collapsed phase with respect to the uncollapsed phase. Instead, we do not detect signatures of superconducting transitions in the specific heat data measured down to 2.5 K, as shown in the following section 3.4.

In order to get clues on the electron-phonon coupling, which is the focus of this study, we fitted the normal state portions of the resistivity curves with a generalized Bloch-Grüneisen law, typical of metals [16,17]:

$$\rho(T) = \rho_0 + \rho_{ph}(T) \quad \text{with} \quad \rho_{ph}(T) = (m-1)\rho'\Theta_D \left(\frac{T}{\Theta_D}\right)^m \int_0^{\Theta_D/T} \frac{z^m}{(1-e^{-z})(e^z-1)} dz \quad (4)$$

where m~3 and $\Theta_D$ is the Debye temperature. Similar values of $\Theta_D \approx 140$ K are obtained for the two phases, consistently with our specific heat data analysis (see section 3.4). However, while for the uncollapsed phase, the experimental resistivity curve departs from the Bloch-Grüneisen fit at high temperature, due to the Ioffe-Regel saturation occurring when the mean free path approaches the lattice spacing, this trend is not apparent for the collapsed phase. The less evident Ioffe-Regel saturation at high temperature in the collapsed phase could point to a weaker effect of the electron-phonon coupling in this phase, however any saturation tendency in this phase could be masked by the lower temperature dependence of resistivity, evidenced by the smaller RRR. Hence, we conclude that the comparison of resistivity curves does not give definitive indications on different roles of phonons in the two phases.

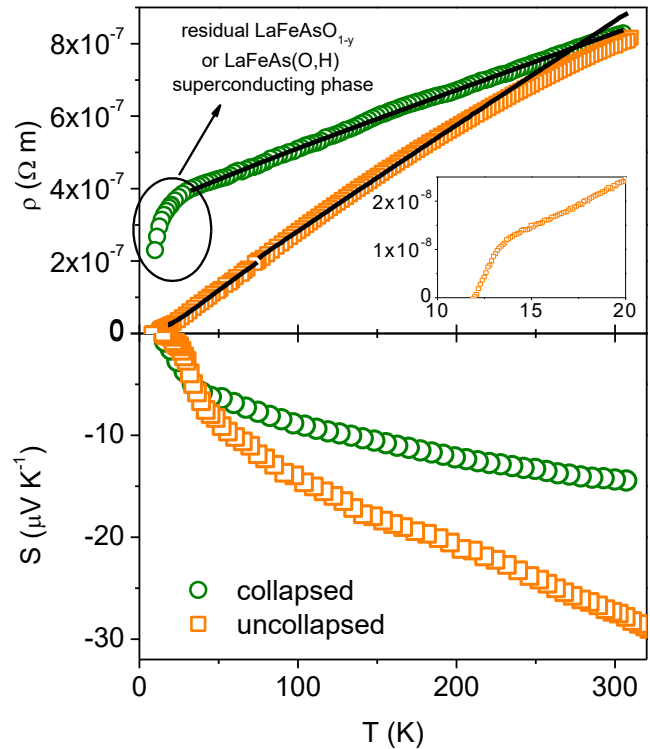

**Figure 2:** Upper panel: resistivity of collapsed and uncollapsed samples as a function of temperature. Fitting with the generalized Bloch-Grüneisen law are displayed as continuous lines. Inset: zoom of the superconducting transition of the uncollapsed sample. Lower panel: Seebeck coefficient measured as a function of temperature in the collapsed and uncollapsed samples.

Magnetotransport characterization was complemented with measurements in magnetic field, shown in the Supplemetary Material file. Magnetoresistivity $(\rho(H)-\rho(H=0))/\rho(H=0)$ is similar in magnitude for the collapsed and uncollapsed phases. The Hall effect is negative and similar in magnitude for the collapsed and uncollapsed phases, exhibiting in both cases non linear field dependence at low temperature T < 100K. The small magnitude and non-linearity of the Hall effect points to the existence of compensated electron and hole bands with opposite contributions to transport.

The Seebeck coefficients, shown in the lower panel of Figure 2, are negative for both phases, consistent with the sign of the Hall effect. The magnitude of Seebeck coefficient of the collapsed phase is about half that of the uncollapsed phase, while being them roughly similar in shape. In both phases, there is a phonon drag bump around 40-50K, which typically accounts for a sizeable electron-phonon interaction, but differently from the uncollapsed phase, in the collapsed phase a $S \sim T^3$ regime at the lowest temperatures is not clearly identified.

The comparison of resistivity (larger in the collapsed phase), magnetoresistivity (similar in the two phases), Hall effect (similar in the two phases) and Seebeck effect (larger in the collapsed phase) could be described assuming that the collapsed phase has higher impurity scattering and lower effective mass. However, as in the case of resistivity, magnetoelectric and thermoelectric properties on their own do not give conclusive indications on differences in electron-phonon coupling in the two phases. In order to investigate this aspect, a direct measurement of phonon spectra is necessary, which is presented in the next section.

**3.3 Raman spectroscopy**

We know from symmetry considerations that four Raman-active phonons are expected for $LaFe_2As_2$ crystals (in both the uncollapsed and collapsed phases) with space group I4/mmm [18]: the $A_{1g}$ mode, in which As atoms are displaced along the c-axis, the $B_{1g}$ mode, in which Fe atoms are displaced along the c-axis, and two $E_g$ modes in which either Fe atoms or As atoms are displaced along the a,b axes (see Figure 4, lower panels). Among them, only $A_{1g}$ and $B_{1g}$ are visible when the polarization of the incident radiation lies in the crystal plane identified by the lattice parameters a,b, while the $E_g$ peaks can appear in the spectrum when the ab plane is tilted. In the present case, a close inspection of the sample surface reveals, in both the uncollapsed and collapsed phases, a rather inhomogeneous structure composed of micro-crystals randomly oriented. Raman measurements performed at ambient conditions at different points of the samples mainly fall into two distinct groups: (i) micro-crystals with a flat and highly reflective surface display two peaks at ~180 and ~190 cm$^{-1}$ and can thus be identified as oriented with the ab plane parallel to the polarization of the incident radiation (*planar configuration*), (ii) micro-crystals with *striped* surface display four peaks at ~110, 180, 190 and 255 cm$^{-1}$ and are thus reasonably oriented in a *tilted configuration* in which all the Raman-active modes are visible. Raman spectra (and images) of uncollapsed crystals belonging to the planar and tilted configuration are shown in Figure 3, panels a and b, respectively, in the 100-265 cm$^{-1}$ range.

Figure 3c displays the Raman spectra of uncollapsed $LaFe_2As_2$ collected in the planar configuration rotating the polarization angle θ in the ab plane. As shown in Figure 3a, each spectrum can be fitted by a two-peak curve: a gaussian centred at ~180 cm$^{-1}$, whose parameters are independent on θ, and a Lorentian centred at ~190 cm$^{-1}$, whose intensity displays a sinusoidal dependence on θ with period π/2, as reported in Figure 3d. Knowing the Raman tensor associated with each mode, the crystal orientation and the polarization vector of the incident and scattered radiation, it is possible to calculate how the intensity of each mode varies with θ, assigning the constant peak at ~180 cm$^{-1}$ to the $A_{1g}$ phonon and the oscillating peak at ~190 cm$^{-1}$ to the $B_{1g}$ phonon; for further details see the Supplementary Material. It is worth noticing that the assignation of the Raman peaks we carried out for $LaFe_2As_2$ shares strong similarities with that reported for parent compounds like $CaFe_2As_2$ [19] and $SrFe_2As_2$ [18], since the vibrational modes we considered mainly involve the motion of Fe and As atoms, identical in the three compounds.

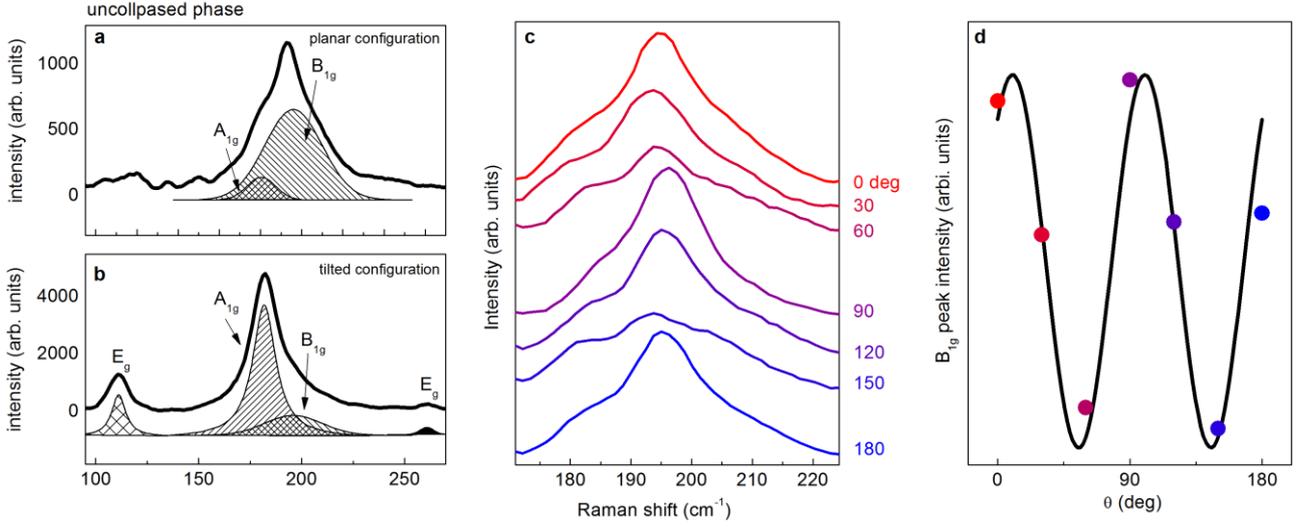

**Figure 3:** a,b) Raman spectra of uncollapsed sample collected in the planar and tilted configurations respectively at ambient conditions. Solid black lines represent the experimental data, while the underlying patterned curves represent the fitting lines to the phonon peaks. c) Polarization-dependent Raman spectra of uncollapsed sample collected in the planar configuration. d) Intensity of the $B_{1g}$ peak at ~190 cm$^{-1}$ as a function of the polarization angle $\theta$. The black curve represents the fitting function $|c\,cos^2(\theta + \theta_0) - c\,sin^2(\theta + \theta_0)|^2$, with free parameters d, $\theta_0$, describing the $\theta$-dependence of the $B_{1g}$ mode once fixed the sample orientation and the polarization direction of the incident and scattered radiation (see Supplementary Material).

Figure 4 (upper panels) shows a direct comparison between the Raman peaks of uncollapsed and collapsed phases in both planar and tilted configurations. Polarization-dependent measurements performed on the collapsed phase, reported in the Supplementary Material file, confirm an assignation of the peaks equivalent to that obtained for the uncollapsed crystal. Regarding the peak positions, we can notice that the frequencies of the $A_{1g}$ mode at ~180 cm$^{-1}$ and the $E_g$ mode at ~255 cm$^{-1}$, which are modes mainly involving the motion of As atoms, remain nearly unchanged in the two phases, while the $B_{1g}$ mode and the low-frequency $E_g$ mode, which are modes mainly involving the motion of Fe atoms, undergo a ~6 cm$^{-1}$ redshift and a ~7 cm$^{-1}$ blueshift respectively going from the uncollapsed to the collapsed phase.

It is interesting to compare our results for $A_{1g}$ and $B_{1g}$, which are modes exclusively involving As and Fe atoms, respectively, with those obtained in the literature for parents compounds, such as SrFe$_2$As$_2$ [20] and NaFe$_2$As$_2$ [21], in pressure-driven uncollapsed-to-collapsed transitions. In ref. [21], the authors observe that the interlayer As-As distance decreases abruptly with pressure, reaching a value that is very close to the As-As covalent bond distance in the collapsed phase. Consequently, they propose that the high-pressure transition from the uncollapsed to the collapsed phase drives the formation of a direct As-As interlayer bond. As for the $B_{1g}$ mode, the authors report a change in the sign of its Grüneisen parameter across the phase transition (from positive to negative), indicating a weakening of the bonds involved in the vibration. In the present case, the situation is quite different, at least from a quantitative point of view. If we compare the changes in some relevant lattice-related quantities across the phase transition at ambient pressure in LaFe$_2$As$_2$, we see that the reduction in the As-As interlayer distance is more than one order of magnitude smaller compared to the pressure-driven transitions, while the Fe-As intralayer distance and the As-Fe-As angles undergo variations that are comparable to that obtained under pressure [1]. It is, thus, well explainable why, in the present case, the passage from the uncollapsed to the collapsed phase affects the Fe-related vibrational mode, while the frequency of the As-related phonon remains practically unchanged.

Since in LaFe$_2$As$_2$ d-orbitals of Fe dominate the composition of the Fermi surface [22], it cannot be ruled out that the topological changes affecting the latter when passing from the collapsed to the superconducting uncollapsed phase might result in a variation of the electron-phonon coupling mechanism in the lattice vibrations involving Fe atoms. Yet, the observed changes in the frequencies of the Fe vibrational modes are not dramatic, and the related effect on the electron-phonon coupling, if any, is likely to be minor.

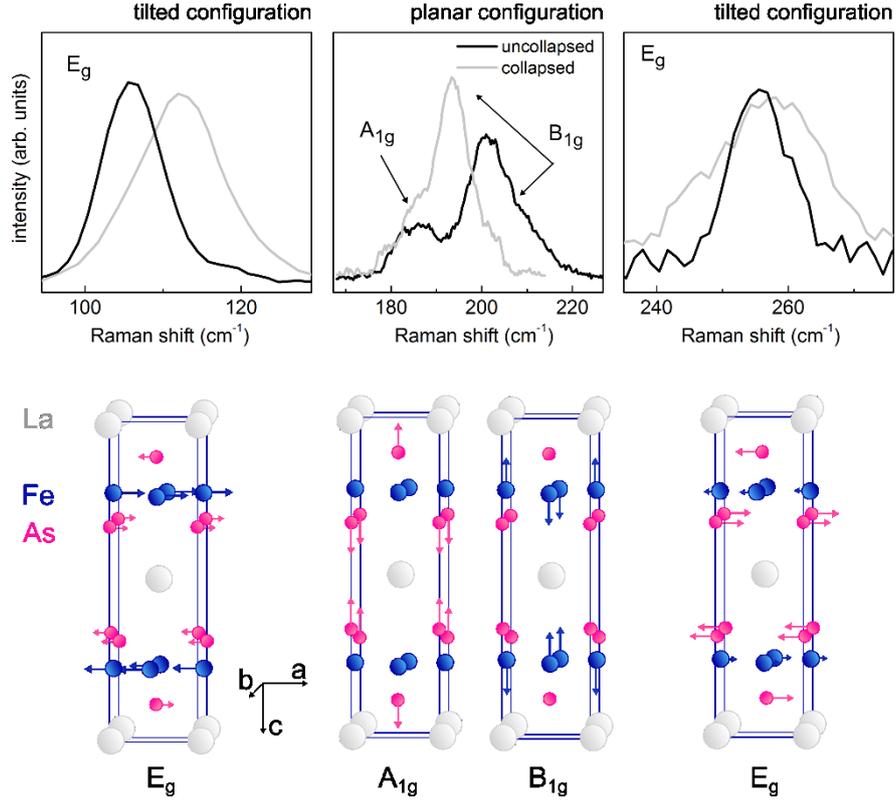

**Figure 4:** Upper panels: comparison between the Raman peaks of uncollapsed (black traces) and collapsed (light grey) traces) samples at ambient conditions. For each mode, we report the spectra collected in the configuration (planar or tilted) that better allows visualising the difference between the two phases. The spectra shown in the upper middle panel are collected with a 1800 lines/mm diffraction grating to better resolve the $A_{1g}$ and $B_{1g}$ peaks. Lower panels: representation of the four Raman-active optical modes in $LaFe_2As_2$.

### 3.4 Specific heat

In Figure 5 (left panel), we present the temperature dependence of the specific heat *c* of both collapsed and uncollapsed samples, normalized to the universal gas constant R, from 2.8 K to 300 K. The overall trend is very similar for the two samples. At higher temperatures the curves tend to saturate to a constant value slightly larger than 15 R, as expected from the Dulong–Petit law for the lattice contribution 3 x $N_{at}$ x R where $N_{at}$=5 is the number of atoms per unit cell. On the other hand, in the low temperature regime, shown in the inset of Figure 5, the electronic contribution is dominant. Here, data are plotted as *c*/T versus $T^2$, in the temperature range below 20 K. The data of the uncollapsed phase exhibit the distinctive bump in correspondence of the superconducting transition, as previously reported [9], while this feature is not observed in the collapsed phase. Being the specific heat a bulk property, the lack of any superconducting feature in the collapsed phase indicates that the amount of superconducting phases present in this sample is small. In order to compare normal state data of the two phases, for the superconducting uncollapsed phase, data in $\mu_0H$= 7 T magnetic field, larger than the upper critical field $H_{c2}(T)$ [9], are also shown. The observed linear trend of normal state data reflects the law:

$$c(T) = \gamma T + \beta T^3 \qquad (1)$$

where $\gamma$ is the Sommerfeld coefficient. In eq. (1), the first term is the electronic specific heat, while the second term $\propto T^3$ is the Debye contribution of acoustic phonons. A linear fit of each set of the data in the inset of Figure 5 give a reasonably good estimation of the Sommerfeld coefficient and the Debye temperature $\Theta_D$, related to the second term by $\Theta_D = \sqrt[3]{\left(\frac{12}{5}\pi^4 R\right)/\beta}$. However, for a more consistent quantitative analysis, the whole curve up to room temperature must be considered.

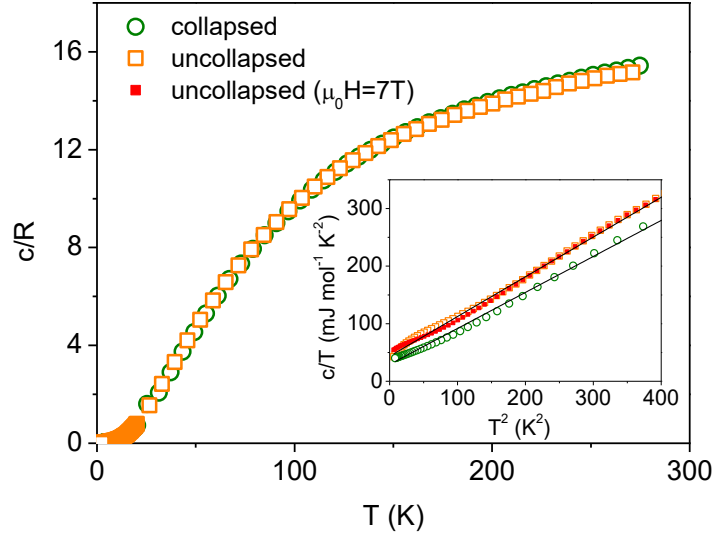

**Figure 5:** Specific heat *c* of collapsed and uncollapsed samples, normalized to the universal gas constant R. Inset: specific heat data plotted as *c*/T versus $T^2$ in the low temperature range below 20 K, with linear fits plotted as black lines. For the superconducting uncollapsed phase, normal state data measured in magnetic field $\mu_0 H$= 7 T, larger than the upper critical field $H_{c2}(T)$, are also shown in the inset.

We assume that the specific heat is a sum of electronic, Debye and Einstein contributions, the latter contributions coming from acoustic and optical phonon modes, respectively:

$$C = C_{el} + C_{Debye} + \sum_i C_{Einstein}^{(i)} \tag{2}$$

$$C_{el} = \gamma T \tag{2.1}$$

$$C_{Debye} = 9 \times R \times \left(\frac{T}{\theta_D}\right)^3 \int_0^{\frac{\theta_D}{T}} \frac{x^4 e^x}{(e^x - 1)^2} dx \tag{2.2}$$

$$C_{opt}^{(i)} = 3 \times R \times \left(\frac{\theta_{E(i)}}{T}\right)^2 \frac{e^{\left(\frac{\theta_{E(i)}}{T}\right)}}{\left(e^{\left(\frac{\theta_{E(i)}}{T}\right)} - 1\right)^2} \tag{2.3}$$

Here, $\Theta_{E(i)}$ are frequencies of optical modes. On the whole, we expect (3 $N_{at}$ - 3) optical modes and 3 $N_{at}$ acoustic modes. In the simplifying assumption that vibrational modes of an ion in the 3 spatial directions are degenerate, eq.s (2.2) and (2.3) are written including one frequency of an acoustic mode $\Theta_D$ and 4 different frequencies of optical modes $\Theta_{E(i)}$, (*i*=1, 2, 3 and 4), each frequency with a three-fold weight. The optical modes are expected to include the Raman-active ones related to ionic displacements in the FeAs planes, depicted in Figure 4, and IR-active mode, all featuring also in other 122 iron pnictides [18,24].

We fitted the experimental data of the two samples with eq.s (2), with fitting parameters $\gamma$, $\Theta_D$, and $\Theta_{E(i)}$ for each sample. In the case of the superconducting uncollapsed phase, data in $\mu_0 H$= 7 T are considered below 20 K, while data in zero field are considered at higher temperature, being the difference between in-field and zero-field data negligible at higher temperature. In Table I, we report the values of acoustic and optical mode frequencies, as well as the Sommerfield coefficients $\gamma$, extracted from the fitting procedure.

**Table I.** Values of $\gamma$, $\Theta_D$, and $\Theta_{E(i)}$ obtained by fitting the normal state specific heat of collapsed and uncollapsed samples. Mode frequency values of the optical modes are reported both in units K and cm$^{-1}$, the latter for easy

comparison with Raman analysis. In the last column, the difference in % between the fitting parameters obtained for the two phases.

| Parameter | collapsed | | uncollapsed | | diff (%) |
|---|---|---|---|---|---|
| $\Theta_{E(1)}$ | 135 (K) | 94 (cm$^{-1}$) | 137 (K) | 95 (cm$^{-1}$) | 1.6 |
| $\Theta_{E(2)}$ | 256 (K) | 178 (cm$^{-1}$) | 271* (K) | 188* (cm$^{-1}$) | 5.5 |
| $\Theta_{E(3)}$ | 344* (K) | 239* (cm$^{-1}$) | 495 (K) | 344 (cm$^{-1}$) | 36.1 |
| $\Theta_D$ | 155 (K) | 108 (cm$^{-1}$) | 148 (K) | 103 (cm$^{-1}$) | -4.8 |
| $\gamma$ | 34 (mJ/mol K$^2$) ** | | 49 (mJ/mol K$^2$) ** | | 36.5 |

* degenerate
** Here the values of $\gamma$ are extracted from the curve fitting in the whole temperature range up to room temperature using eq.s (2) and they are larger than the values obtained by the fit of low temperature data using eq. (1), as performed in ref. [9] for the uncollapsed phase and also displayed in the inset of Fig. 5 for the two phases.

From the best fit, the collapsed phase has twice degenerate $\Theta_{E(3)}$ mode, while the uncollapsed phase has twice degenerate $\Theta_{E(2)}$ mode, however the quality of the fit changes very little if $\Theta_{E(3)}$ is assumed twice degenerate for the uncollapsed phase or $\Theta_{E(2)}$ is assumed twice degenerate for the collapsed phase. Hence, we can say that the specific heat data are sensitive to 3 optical modes, and one acoustic mode. As reported in Table I, the optical modes of the collapsed phase have a frequencies 94 cm$^{-1}$, 178 cm$^{-1}$, and 239 cm$^{-1}$, and the acoustic mode 108 cm$^{-1}$ in the collapsed phase. Comparing the relative differences of the two phases, the optical modes with lower frequencies $\Theta_{E(1)}$ and $\Theta_{E(2)}$ differ by few percent between the two phases, while the higher frequency modes $\Theta_{E(3)}$ is significantly higher for the uncollapsed phase. However, the $\Theta_{E(3)}$ values suffer of larger indeterminacy because their relative contribution to the specific heat is negligible in most of the considered temperature range except at the highest temperatures (above ~ 100 K); consequently the difference of $\Theta_{E(3)}$ values in the two phases is not reliable. Finally, the acoustic modes, represented by the Debye parameter $\Theta_D$, shifts by ~5% to higher energies from the uncollapsed to the collapsed phase.

In general, small frequency shifts of the phonon modes are hardly appreciable in the specific heat analysis. More interesting and significant result of the specific heat analysis is that the Sommerfield coefficients differ by around 36% between the two phases. We note that the Sommerfield coefficient can be expressed as:

$$\gamma = \frac{\pi^2 K_B^2 N_{av}}{2} \frac{1}{E_F} \qquad (3)$$

where $K_B$ is the Boltzmann constant, $N_{av}$ is the Avogadro number and $E_F$ is the Fermi energy, which for three-dimensional parabolic bands can be expressed in terms of the effective mass $m_{eff}$ as $E_F = \frac{\hbar^2}{2m_{eff}}(3\pi^2 n)^{\frac{2}{3}}$. Therefore, the significantly larger value of $\gamma$ extracted for the uncollapsed phase points to a larger effective mass in this phase. Note that even the small amount of LaFeAsO impurity phase in the collapsed sample (see Figure 1) may contribute to the value of $\gamma$. However, considering a Sommerfield coefficient $\gamma \approx 26$ measured on LaFeAsO [23], even assuming that the amount of impurity LaFeAsO is as large as 10%, we would extract a $\gamma$ value of ~35 mJ/mol K$^2$ instead of 34 mJ/mol K$^2$ for the collapsed phase, in any case much smaller than the value 49 mJ/mol K$^2$ extracted for the uncollapsed sample, which could be itself similarly underestimated, due to the presence of metallic impurity phases LaFeAsO, LaAs, and FeAs in the uncollapsed sample. The larger Seebeck effect in the uncollapsed phase (see Figure 2) is also consistent with larger effective mass in this phase. The larger effective mass reflects the importance of correlations predicted by theoretical works [8], which also represent a key ingredient for the appearance of superconductivity.

## 4. Discussion and conclusions

We start with comparing the frequency values of phonon modes obtained from the specific heat fit with the frequencies of Raman peaks shown in the previous sections. It is natural to identify the specific heat middle mode $\Theta_{E(2)}$ with the contributions of both the out-of-plane Raman modes A$_{1g}$ at ~180 cm$^{-1}$ and B$_{1g}$ at

~190 cm$^{-1}$. The lowest frequency mode $\Theta_{E(1)}$ mode could be identified with the low-frequency in-plane $E_g$ mode, and it could also include the contribution of the IR low energy $A_{2u}$ mode, which is expected for the 122 compounds at ~100 cm$^{-1}$ [24,20]. The identification of the highest $\Theta_{E(3)}$ modes is more ambiguous, because, as already mentioned, their estimation affected by larger indeterminacy. $\Theta_{E(3)}$ could include the contribution of the high-frequency optical modes, namely the in-plane $E_g$ mode at ~255 cm$^{-1}$ observed with Raman spectroscopy and the IR-active modes $A_{2u}$ and $E_u$, which are expected in the range of 280-320 cm$^{-1}$ in 122 compounds [18,20]. Finally, the Debye parameter $\Theta_D$, which represents acoustic phonons, may be identified with the acoustic low energy $E_u$ mode.

In order to draw information of possible differences in phonon spectra between the collapsed and uncollapsed phases, we compare the results obtained from the above characterizations.

In the Raman analysis, the out-of-plane Raman modes $A_{1g}$ at ~180 cm$^{-1}$ is unchanged between the collapsed and uncollapsed phases while $B_{1g}$ at ~190 cm$^{-1}$ undergoes a ~6 cm$^{-1}$ redshift from the uncollapsed to the collapsed phase. Consistently, in the specific heat analysis, the $\Theta_{E(2)}$ mode is redshifted by ~5% from the uncollapsed to the collapsed phase. The lowest frequency $E_g$ mode undergoes a ~7 cm$^{-1}$ blueshift in the Raman analysis from the uncollapsed to the collapsed phase; however no appreciable change is found for the specific heat $\Theta_{E(1)}$ between the two phases.

Furthermore, the acoustic mode $\Theta_D$, is found to shift to higher energies from the uncollapsed to the collapsed phase, and this is consistent with theoretical calculations that predict a stiffening of the non Raman-active acoustic $A_{2u}$ mode in other 122 pnictide compounds that exhibit an uncollapsed-to-collapsed phase transition with applied pressure [20,21].

Gathering the key findings obtained from our experimental characterizations, and we draw two main conclusions:

1) The phonon spectrum slightly changes between the two phases, yet not quite significantly. If a strong electron-phonon coupling was at work in the uncollapsed superconducting phase, we would expect a significant broadening of the width of the Raman peak involved in the coupling mechanism (e.g. as that observed in the $E_{2g}$ phonon of $MgB_2$ compared to the isostructural crystal $AlB_2$ [25,26]). In the present case, instead, the $A_{1g}$ and $B_{1g}$ peaks show comparable widths in the two phases, while the $E_g$ peaks are slightly broader in the collapsed phase, possibly due to a higher degree of crystalline disorder, as discussed above. In the same vein, no clear evidence of different electron-phonon coupling is obtained from transport data.

2) the Sommerfield constant is significantly different in the two phases, namely 34 (mJ/mol K$^2$) in the collapsed phase and much larger, 49 (mJ/mol K$^2$), in the uncollapsed phase; the uncollapsed phase also exhibits twice as large Seebeck coefficient with respect to the collapsed phase; both these findings indicate much higher electron correlation in the superconducting uncollapsed phase.

Our previous experimental study in the uncollapsed phase [9] evidenced a high effective mass and strong electron-phonon coupling, the latter possibly playing a role in the appearance of superconductivity. With the comparison between the superconducting uncollapsed phase and the non-superconducting collapsed phase carried out in the present study, we conclude that the phonon spectrum is unlikely to play a significant role in determining the appearance of superconductivity in the uncollapsed phase. On the other hand, strong electron correlation in the uncollapsed phase likely plays a primary role, as predicted by theoretical works [4,5,8]. Indeed, according to theory [8], in uncollapsed $LaFe_2As_2$ correlations are even stronger than in other 122 pnictides such as $BaFe_2As_2$ when compared at the same nominal valence of electrons per unit cell, mainly due to the smaller bare bandwidth of $LaFe_2As_2$.

# Supplementary Material

**Polarization dependent Raman measurements**

The experimental configuration adopted for the polarization-dependent Raman measurements consists of a λ/2 waveplate to rotate the polarization on the sample surface and a polarization analyser. The former was placed along the laser optical path, behind the microscope objective, while the latter was located before the diffraction grating to select the vertical component of the scattered light. In this configuration, the intensity I of a mode described by the Raman tensor A is:

$$I_A \propto |\varepsilon_{in} R^T(\theta) A R(\theta) \varepsilon_{out}|^2 \qquad (S1)$$

where $\varepsilon_{in}$ is the polarization vector of the laser beam, $R(\theta)$ is the rotation matrix relative to the propagation direction of the beam, and $\varepsilon_{out}$ is the polarization of the detected radiation. In the present experiment, the analyzer was oriented so that $\varepsilon_{in} \| \varepsilon_{out}$. We choose the reference system so that x∥a, y∥b, z∥c (with a,b,c lattice parameters). Therefore, in the planar configuration, in which the crystal plane identified by the a,b parameters is perpendicular to the propagation direction of the incident light,

- the λ/2 waveplate rotates the polarization by an angle θ about the z-axis: $R(\theta) = R_z(\theta)$
- the polarization vector of the laser beam is parallel to the x-axis: $\varepsilon_{in}$ = (1 0 0)
- the polarization vector of the scattered radiation is selected along the x-axis: $\varepsilon_{out}$ = (1 0 0)$^T$

Applying eq. (S1) to the Raman tensors associated with the Raman-active modes $A_{1g}$, $B_{1g}$, $1E_g$, $2E_g$

$$\begin{pmatrix} a & 0 & 0 \\ 0 & a & 0 \\ 0 & 0 & b \end{pmatrix} \quad \begin{pmatrix} c & 0 & 0 \\ 0 & -c & 0 \\ 0 & 0 & 0 \end{pmatrix} \quad \begin{pmatrix} 0 & 0 & 0 \\ 0 & 0 & -c \\ 0 & -e & 0 \end{pmatrix} \quad \begin{pmatrix} -e & 0 & 0 \\ 0 & 0 & e \\ 0 & 0 & e \end{pmatrix}$$

we find that:

- $I_{A_{1g}} = constant$
- $I_{B_{1g}} = |c \, cos^2(\theta) - c \, sin^2(\theta)|^2$
- $I_{1E_g} = 0$
- $I_{2E_g} = 0$

In the following, we report the results of polarization-dependent Raman measurements in the collapsed phase, which behave in perfect analogy with that performed on the uncollapsed phase shown in the main text.

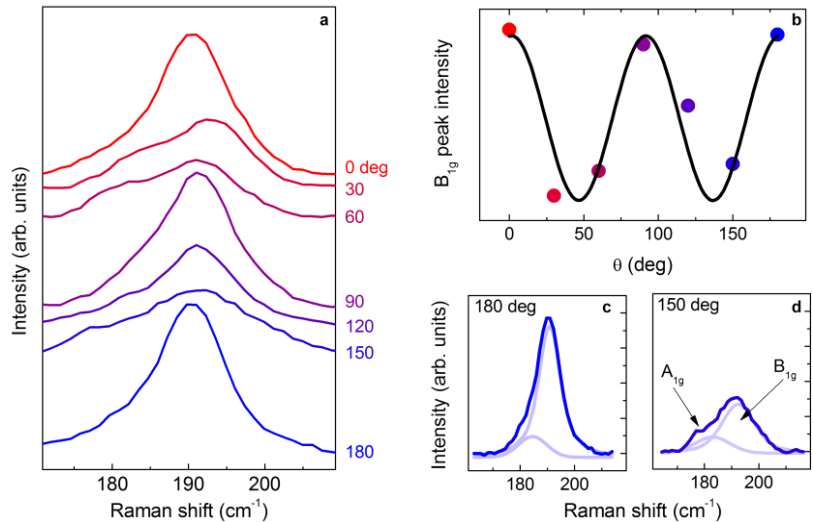

**Figure S1:** . a) Polarization-dependent Raman spectra of collapsed sample collected in the planar configuration. b) Intensity of the $B_{1g}$ peak at ~190 cm$^{-1}$ as a function of the polarization angle θ. The black curve represents the fitting function $|c \, cos^2(\theta + \theta_0) - c \, sin^2(\theta + \theta_0)|^2$, with free parameters c, $\theta_0$, describing the θ-dependence of the $B_{1g}$ mode once fixed the sample orientation and the polarization direction of the incident and scattered. c,d) Examples of fit of the Raman peaks in the planar configuration, where the $A_{1g}$ fitting parameters remain practically constant on varying the polarization angle θ, while the intensity of the $B_{1g}$ follows the trend reported in panel b.

**Magnetotransport measurements**

Magnetoresistivity, shown in Figure S2, is similar in magnitude for the collapsed and uncollapsed phases.

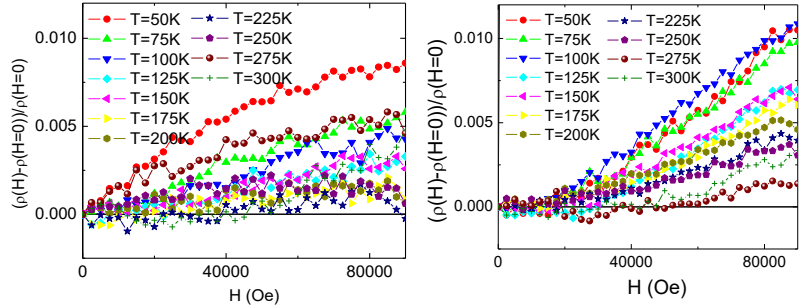

**Figure S2:** Magnetoresistivity (ρ(H)-(ρ(H=0))/(ρ(H=0) plotted as a function of magnetic field at different temperatures in the collapsed (left panel) and uncollapsed (right panel) samples.

The Hall effect, shown in Figure S3, is similar in sign (negative), magnitude, and non-linear field dependence at low temperature T < 100K for the collapsed and uncollapsed phases. The small magnitude and non-linear field dependence of the Hall effect reflect the compensation of electrons and holes in these materials and their multiband character gives rise to different temperature dependence of the Hall voltage versus field slope, which slightly decreases (increases) with increasing temperature in the collapsed (uncollapsed) phase. The non-linear dependence of the Hall voltage on the magnetic field could also originate from charge scattering with magnetic excitations.

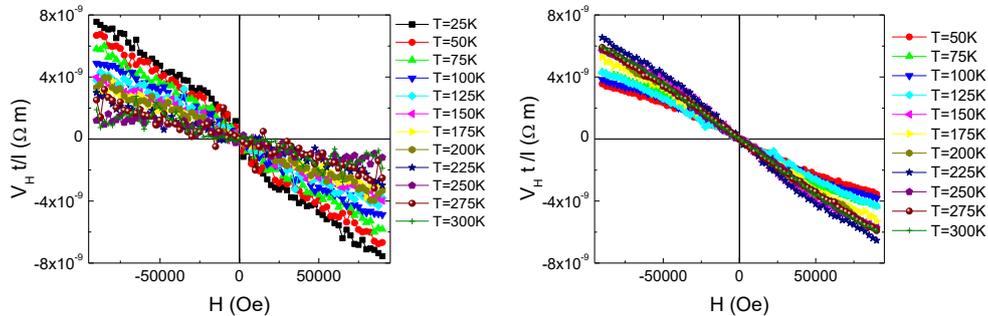

**Figure S3:** Transverse resistance $R_H=V_H/I$, multiplied by thickness t, plotted as a function of magnetic field at different temperatures in the collapsed (left panel) and uncollapsed (right panel) samples.